\documentclass[fleqn,usenatbib,useAMS]{mnras}

\usepackage{xurl}
\usepackage{booktabs}
\usepackage{tabularx}
\usepackage{hyperref}
\usepackage[multiple]{footmisc}
\usepackage{newtxtext,newtxmath}
\usepackage{graphicx}

\newcommand{\UCB}{Berkeley SETI Research Center, University of California, Berkeley, CA 94720, USA}
\newcommand{\seti}{SETI Institute, 339 Bernardo Ave, Suite 200, Mountain View, CA 94043, USA}
\newcommand{\oxford}{Breakthrough Listen, University of Oxford, Department of Physics, Denys Wilkinson Building, Keble Road, Oxford, OX1 3RH, UK}
\newcommand{\UNM}{Department of Physics and Astronomy, University of New Mexico, Albuquerque, NM 87131, USA}
\newcommand{\SARAO}{South African Radio Astronomy Observatory, Liesbeek House, River Park, Gloucester Road, Mowbray, Cape Town, 7700, South Africa}
\newcommand{\orleans}{LPC2E, OSUC, Univ Orleans, CNRS, CNES, Observatoire de Paris, F-45071 Orleans, France}
\newcommand{\ska}{SKA Observatory, 26 Dick Perry Avenue, Kensington, WA 6151, Australia}
\newcommand{\malta}{University of Malta, Institute of Space Sciences and Astronomy, Malta}
\newcommand{\binitiatives}{Breakthrough Initiatives, Moffett Field, CA, USA}

\title[BL's Technosignature Survey with MeerKAT]{Breakthrough Listen's Automated Commensal Technosignature Survey with MeerKAT}

\author[D. J. Czech et al.]{Daniel J. Czech,$^{1}$\thanks{Contact e-mail: daniel.czech@physics.ox.ac.uk}
David H.E. MacMahon,$^{2}$
Ian Heywood,$^{1}$
Chenoa D. Tremblay,$^{2,3,4}$
Matt Lebofsky,$^{2}$
\newauthor Kevin Lacker,$^{2}$
Cherry Ng,$^{5}$
Dave Horn,$^{6}$
Sarah Buchner,$^{6}$
Brian Lacki,$^{2}$
Alex Andersson,$^{1}$
\newauthor Joe S. Bright,$^{3}$
Steve Croft,$^{1,2,3}$
Dave R. DeBoer,$^{2}$
Jamie Drew,$^{9}$
Vishal Gajjar,$^{2,3}$
Peter Ma,$^{2}$
\newauthor Alex W. Pollak,$^{3}$
Danny Price,$^{1,7}$
Mark Ruzindana,$^{2}$
Andrew P.V. Siemion,$^{1,2,3,8}$
S. Pete Worden,$^{9}$
\newauthor Fernando Camilo$^{6}$
\\
$^{1}$\oxford\\
$^{2}$\UCB\\
$^{3}$\seti\\
$^{4}$\UNM\\
$^{5}$\orleans\\
$^{6}$\SARAO\\
$^{7}$\ska\\
$^{8}$\malta\\
$^{9}$\binitiatives
}

\begin{document}
\label{firstpage}
\pagerange{\pageref{firstpage}--\pageref{lastpage}}
\maketitle

\begin{abstract}
 
The search for extraterrestrial intelligence (SETI) is an ongoing effort to detect technosignatures, evidence of technologically capable life beyond Earth. Conducting a comprehensive SETI programme requires a large amount of telescope time, which must be balanced with the science goals of a given observatory. Fortunately, many modern radio telescopes offer commensal access to the data they produce, allowing multiple scientific programmes to operate in parallel. The MeerKAT radio telescope in South Africa provides commensal access to a range of components, from each antenna's digitiser to the main channeliser (F-engine), via multicast Ethernet groups. Here, we describe the Breakthrough Listen user-supplied equipment (BLUSE) system at MeerKAT, which leverages multicast Ethernet to conduct an autonomous commensal technosignature survey, processing the full available bandwidth from all antennas. Its primary mode of operation is to upchannelise the incoming F-engine data to $\sim1$Hz resolution, synthesize coherent beams on objects of interest, and search the resultant data for technosignatures. Since 2022, BLUSE has autonomously processed data from coherent beams synthesized on more than 1.2 million individual pointings, including repeat visits. BLUSE demonstrates how commensal technosignature surveys on radio telescope arrays offer a rapid and cost-effective way to increase the rate at which technosignature surveys can be conducted. This article describes the architecture of BLUSE, provides experimental evidence validating its features and performance, and quantifies its observing progress over the past few years. We also discuss the technical evolution of BLUSE, examine challenges faced and addressed, and consider avenues for future research and development.  

\end{abstract}


\section{Introduction} 
\label{sec:intro}

Until relatively recently, most technosignature searches were conducted as targeted surveys with sensitive single-dish telescopes (e.g. \citealt{Tarter_1980}, \citealt{Horowitz_1993}, \citealt{Siemion_2013}, and \citealt{Enriquez_2017}). Such telescopes tend to be large, with a small primary field of view, and must in most cases be pointed at targets mechanically, one by one. In prior technosignature surveys with single dishes, sky localisation (and radio frequency interference rejection) has frequently been performed by means of a cadence-based observing strategy \citep{Sheikh_2020, Choza_2023, Margot_2023} where observations of the target are interspersed with other off-source targets. The appearance of a signal in on-source observations but not in off-source observations is evidence that it may be localised to the target. However, such an approach is time-inefficient and susceptible to radio frequency interference (entering, for example, via the sidelobes of the antenna) which may be interpreted as a potential sky-localised signal. Single-dish surveys, even the most recent, tend to observe targets that number in the thousands \citep{Price_2020}. Even when considering the significant ``by-catch'' (objects which, while not the intended target of an observation, are none the less present in the primary field of view of the telescope), the number of catalogued objects in the survey footprint still remains in the tens of thousands \citep{Wlodarczyk-Sroka_2020,Garrett_2023,Margot_2023}.

Array radio telescopes offer many benefits for technosignature surveys, e.g. \cite{Houston_2021}. Their dishes tend to be small (compared to single-dish radio telescopes) and numerous, providing a generous field of view without sacrificing collecting area and sensitivity. The Allen Telescope Array, for example, was originally built to capitalise on these advantages \citep{Welch_2009}. Furthermore, arrays can be used for interferometric imaging and beamforming, providing instantaneous sky localisation and RFI rejection capabilities. Recent technosignature surveys making use of these capabilities have been implemented at the Karl G. Jansky Very Large Array \citep{Tremblay_2024}, the Allen Telescope Array \citep{Farah_2023, Sheikh_2023, Saide_2023} and the Murchison Widefield Array \citep{Tingay_2016, Tingay_2018, Tremblay_2022} among others. VLBI techniques have also been applied to the search for technosignatures with e-MERLIN and the European VLBI Network, for example \citep{Wandia_2023}.

The case for large, target-agnostic technosignature surveys maximising sky and frequency coverage is easy to make when considering attempts to quantify the ``cosmic haystack" search space \citep{Wright_2018}. Arrays constructed with a multicast Ethernet-based architecture, such as MeerKAT \citep{Slabber_2018} and the MWA \citep{Morrison_2023}, or fitted with other means to split the data, such as the VLA \citep{Tremblay_2024}, enable rapid sky coverage at low marginal cost. MeerKAT offers this capability and enables ``User Supplied Equipment" systems to access such data via the SPEAD2\footnote{An implementation of SPEAD (Streaming Protocol for Exchanging Astronomical Data) with both Python and C++ bindings. \href{https://github.com/ska-sa/spead2}{https://github.com/ska-sa/spead2}} protocol \citep{Manley_2010}. 
Breakthrough Listen's User Supplied Equipment system (BLUSE), the subject of this work, is currently observing (commensally) at a rate of approximately 29,000 new (that is, never observed before by BLUSE) stars per month for 290 seconds per star. Clearly, array-based commensal technosignature surveys occupy a critical role in the portfolio of extant projects searching for extraterrestrial intelligence (SETI), driving advances in the field forward with the sheer volume of their results. 
 
\begin{figure*}
    \centering
    \includegraphics[width=\textwidth]{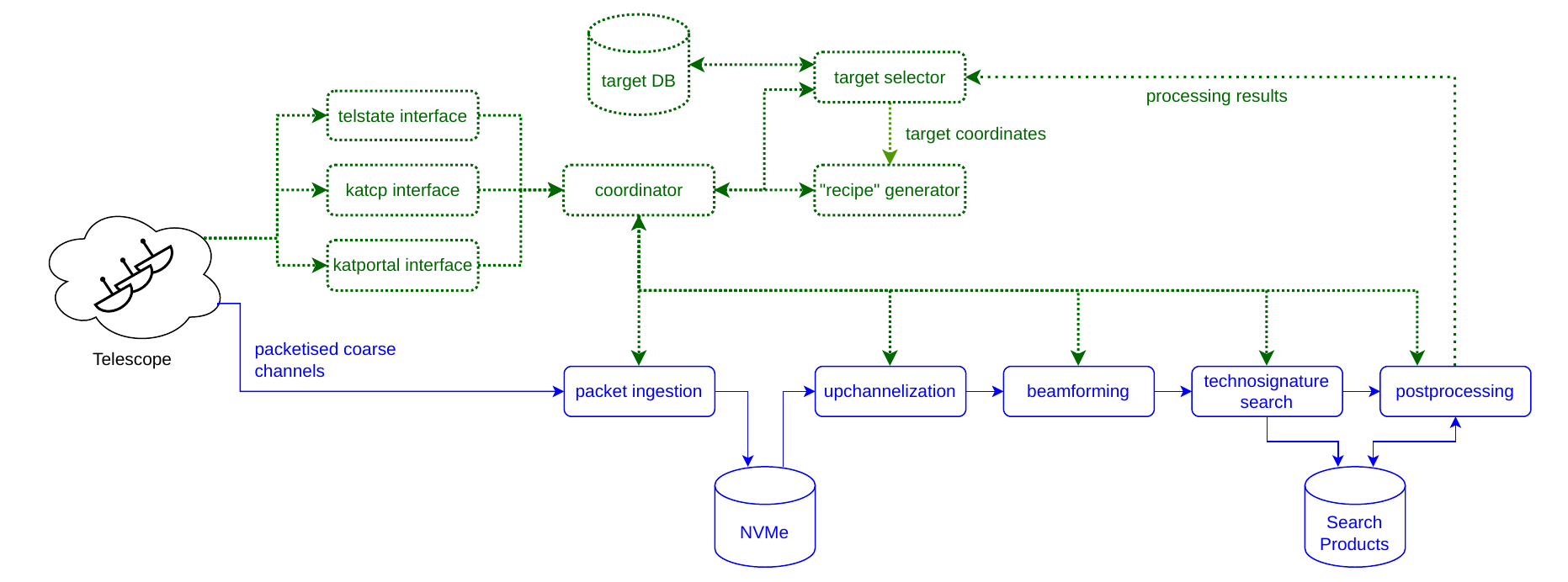}
    \caption{High-level diagram of the BLUSE commensal technosignature survey system. The dotted green lines represent the flow of metadata and control messages, while the solid blue lines represent the high-speed data path.}
    \label{fig:system}
\end{figure*}

BLUSE is an automated commensal beamformer and technosignature search system. Figure \ref{fig:system} provides a high-level outline of the system's architecture: Data are received directly from the MeerKAT F-engines and upchannelised to $\sim1$\,Hz resolution. Beams are synthesized on targets which fall within the primary field of view, and a Taylor-tree de-Doppler drift search \citep{sheikh2019, Cohen_1987} is conducted on each beam. The search products (``hits" and ``stamps"\footnote{\href{https://github.com/lacker/seticore}{https://github.com/lacker/seticore}}, see Section \ref{sec:products}) are then saved to disk. 

The high frequency resolution used by BLUSE enables the detection of extremely narrowband signals, which cannot be produced by known natural astrophysical processes, and their frequency drift due to Doppler acceleration \citep{Li_jk_2022}. For example, one of the most narrowband astrophysical features known is an OH maser line 550\,Hz wide \citep{Cohen_1987}. Narrowband radio signals are also energy efficient to transmit over interstellar distances, relative to other known means of communication. For these reasons, $\sim$Hz-scale resolution has long been a focus of prior SETI surveys (see, for example, \citealt{Horowitz_1993, Siemion_2013, Choza_2023}).

In this work, we introduce the BLUSE data pipeline in Section \ref{sec:pipeline}, observing automation system in Section \ref{sec:automation}, and hardware in Section \ref{sec:hardware}. In Section \ref{sec:validation} we provide thorough validation and evidence of the system's functionality and features. Section \ref{sec:progress} describes the observing progress thus far. Finally, we conclude in Section \ref{sec:conclusion}.

\section{Processing pipeline} 
\label{sec:pipeline}

MeerKAT is a 64-antenna radio interferometer located in the Karoo desert, South Africa \citep{Jonas2016}. It consists of 64 uniform 13.5-m offset Gregorian antennas, each equipped with three receivers operating from 544\,MHz to 3.5\,GHz (see Table \ref{tab:data-rates}). The MeerKAT correlator uses an FX architecture where signal channelisation is performed via a first-stage `F-engine' before pairwise cross-correlations are performed using an `X-engine.' MeerKAT uses a packetized Ethernet architecture to implement a cornerturn, whereby like channels from all antennas are grouped together for cross-correlation by sending a subset of channels in each packet to a destination X-engine \citep[see ][]{Parsons2008,Hickish2018}. MeerKAT makes use of IPv4\footnote{IPv4: Internet Protocol version 4} multicast to allow multiple digital backends to connect to the F-engine data streams simultaneously. A number of user-supplied backends, known as user-supplied equipment (USE), subscribe via multicast to (e.g.) the F-engine data and perform alternative signal processing tasks in parallel with the correlator (e.g. \cite{sanidas2017}, \cite{bailes2020}).

BLUSE follows this methodology, ingesting F-engine data from the multicast network and performing an autonomous technosignature search. Figure \ref{fig:system} provides a high-level outline of the entire system. The high-speed data processing pipeline begins with subscriptions to the appropriate F-engine multicast groups and receiving incoming Ethernet UDP (User Datagram Protocol) packets. MeerKAT offers three receivers (UHF, L- and S-band) and three wide-band observing modes (1k, 4k and 32k coarse channels, set by the primary observer) as well as a narrowband ``zoom" mode\footnote{\url{https://skaafrica.atlassian.net/wiki/spaces/ESDKB/overview}}. We do not make use of the zoom mode since we need to upchannelise across the entire band, and the wide-band modes are always available. The S-band receivers offer more bandwidth than the F-engines can process, so the band is split into five overlapping bands (S0 to S4). The observing modes and their associated parameters are retrieved by the automation and interfacing software, described in Section \ref{sec:automation}. The data rate output by each antenna's F-engine varies by the bandwidth of each observing mode (see Table \ref{tab:data-rates}). At L-band, the data rate produced by a single antenna's F-engine is 856\,MHz $\times$ 8 bits $\times$ 2 (complex) $\times$ 2 polarisations = 27.392\,Gbps, or approximately 1.753\,Tbps for a full array of 64 antennas. The actual data rate that traverses the network is higher due to networking overhead: for every 1024 bytes of data, the Ethernet/IP/UDP packet headers add 42\,bytes and the SPEAD2 header adds 96\,bytes. Therefore, the total data rate to be received by one processing instance is approximately 31.08\,Gbps at L-band when all 64 antennas are in use.

\begin{table*}
\centering
\caption{MeerKAT receivers, modes and associated data rates received by BLUSE per processing instance, and the resultant fine channel bandwidth. BLUSE ingests and processes the coarse channels emitted by the F-engines. Note: S-band is split into 5 overlapping 875 MHz subbands covering 1750--3500 MHz.}
\footnotesize
\begin{tabular}{lllllll}
\toprule
\textbf{Receiver} & \textbf{Digitised} & \textbf{Digitised Freq.} & \textbf{F-engine Mode} & \textbf{Coarse Chan.} & \textbf{Fine Chan.} & \textbf{Gbps/instance}\\
& \textbf{BW [MHz]} & \textbf{Range [MHz]} &  & \textbf{BW [kHz]} & \textbf{BW [Hz]} & \\
\midrule
UHF & 544 & 544--1088 & 1k & 531.25 & 1.01 & 17.408\\
UHF & 544 & 544--1088 & 4k & 132.81 & 1.01 & 17.408\\
UHF & 544 & 544--1088 & 32k & 16.6  & 1.01 & 17.408\\
L & 856 & 856--1712 & 1k & 835.94 & 1.59 & 27.392\\
L & 856 & 856--1712 & 4k & 208.98 & 1.59 & 27.392\\
L & 856 & 856--1712 & 32k & 26.12 & 1.59 & 27.392\\
S & 875 & 1750--3500 (5 subbands) & 1k & 854.49 & 1.62 & 28.0\\
S & 875 & 1750--3500 (5 subbands) & 4k & 213.62 & 1.62 & 28.0\\
S & 875 & 1750--3500 (5 subbands) & 32k & 26.7 & 1.62 & 28.0\\
\bottomrule
\end{tabular}
\label{tab:data-rates}
\end{table*}

Each processing pipeline instance handles an amount of incoming data equivalent to that produced by one MeerKAT antenna. However, the data received by each instance (via the multicast groups) consist of a common segment of bandwidth (1/64$^{th}$) repeated for every antenna. Thus, each instance is responsible for upchannelising, beamforming, and searching 1/64$^{th}$ of the full bandwidth. Table \ref{tab:data-rates} provides the different receiver-specific bandwidths for these segments.

SPEAD2 packets are received by $\texttt{hpguppi\_daq}$ instances\footnote{\href{https://github.com/UCBerkeleySETI/hpguppi\_daq}{https://github.com/UCBerkeleySETI/hpguppi\_daq}}, the payload voltages assembled into Green Bank Ultimate Pulsar Processing Instrument ($GUPPI$) raw format \citep{GUPPI, Lebofsky_2019} blocks and written to non-volatile memory express (NVMe) modules in RAID~0\footnote{RAID 0 stripes data across each disk in the array to improve read/write performance at the expense of redundancy} to manage the incoming data rate. The use of NVMe modules permits much deeper buffers than would be feasible with RAM. We investigated the performance and write endurance of different drives for our specific use case: sequentially writing to the disks to capacity, reading the data back, and deleting all of the data. Using purpose-built software\footnote{\href{https://github.com/david-macmahon/disk\_hammer}{https://github.com/david-macmahon/disk\_hammer}}, we tested various different drives under consideration to failure and found their longevity to be far greater than their stated specifications (for our particular usage pattern, linear writing, reading and deletion).

\subsection{Calibration}
\label{sec:calibration}

In order to coherently combine data streams from antennas to form synthesized beams, a set of phasing solutions is required. BLUSE can retrieve and rely upon calibration solutions produced for the primary observer by MeerKAT's Science Data Processor. The MeerKAT calibration pipelines are discussed extensively in the MeerKAT External Service Desk Knowledge Base\footnote{\url{https://skaafrica.atlassian.net/wiki/spaces/ESDKB/pages/338723406/}}. 

Upon commencing a recording, BLUSE retrieves the latest calibration solutions from {\sc TelState}\footnote{\href{https://github.com/ska-sa/katsdptelstate}{https://github.com/ska-sa/katsdptelstate}} and supplies them to the beamformer for application. See Section~\ref{sec:automation} and Section~\ref{sec:validation} for further details. \\

\subsection{Upchannelising and beamforming}
\label{sec:beamforming}

For each instance, the incoming data are further channelised to approximately 1\,Hz resolution, which we refer to as ``upchannelisation" (see Table \ref{tab:data-rates} for precise values). 

Following upchannelisation, each instance then forms the desired synthesized beams on targets within the primary field of view. BLUSE currently forms 64 coherent beams and one incoherent beam per 290~s primary pointing. The calibration solutions obtained from {\sc TelState} and the coordinates for the 64 targets chosen in the primary field of view are written to an HDF5-based file by the \texttt{bfr5\_generator} process (see Section~\ref{sec:automation-processes}). We refer to this file as a ``beamformer recipe file".  The \texttt{bfr5\_generator} also calculates delays and delay rates for each beam (relative to the boresight pointing, which the F-engines track in delay and phase) at one second intervals over the duration of the recording and writes them into this beamformer recipe file for later access by the beamformer.

See Section~\ref{sec:targets} for a detailed explanation on target selection and Section~\ref{sec:validation} for experimental validation of the beamformer. Both upchannelisation and beamforming are conducted with the \texttt{seticore} software.

\subsection{Technosignature search}
\label{sec:search}

A Taylor-tree narrowband de-Doppler drift search \citep{Taylor_1974, sheikh2019, Siemion_2013} is conducted on the beamformed data, using \texttt{seticore}\footnote{seticore includes a high performance implementation of the search algorithm in TurboSETI, a widely used de-Doppler drift search software package \citep{Enriquez_2017}}. Many technosignature search algorithms have been proposed, for example, searching for narrowband pulsed signals using a fast folding algorithm \citep{Suresh_2023} or using machine learning approaches such as variational autoencoders \citep{Ma_2023}. Taylor-tree approaches, however, have thus far been among the most widely used where very high throughput is required. We anticipate that more sophisticated approaches will be used in tandem with \texttt{seticore} in the near future (see Section~\ref{sec:conclusion}).

Recent searches for narrowband radio technosignatures have searched a variety of drift rate ranges: for example, $\pm 2$\,Hz/s \citep{Enriquez_2017}, $\pm 4$\,Hz/s \citep{Price_2020, Choza_2023}, $\pm 8.86$\,Hz/s \citep{Margot_2023} and $\pm 50$\,Hz/s \citep{Tremblay_2024}. For BLUSE, we use a drift rate range of $\pm 10$\,Hz/s, and intend to expand this to wider values in the future.  \citet{Li_2023} consider the distribution of drift rates that would arise from transmitters on the surfaces of exoplanets. They find that a drift rate of $\pm 44$\,Hz/s would be sufficient to account for 99\% of potential signals. Of course, searching wider drift rate ranges incurs a computational cost which must be balanced against metrics like throughput. 

We used an initial signal-to-noise (SNR) threshold of six, selected to provide a reasonable balance between sensitivity to potential signals of interest and avoidance of spurious detections. Adjusting this threshold directly affects the rate at which candidate signal ``hits" are recorded for further analysis, and thus the rate at which BLUSE consumes available storage space. Therefore, we anticipate adjusting this parameter during operations.

\subsection{Data products}
\label{sec:products}

Always-on commensal technosignature surveys implemented at arrays must be able to manage an enormous incoming data rate (see Table~\ref{tab:data-rates}). At most radio telescope arrays, it is infeasible to record and store raw F-engine voltages from each antenna in perpetuity, due to the storage volume that would be required. By contrast, while not routine, prior single-dish surveys have occasionally been able to do so \citep{Lebofsky_2019} as there are usually far fewer receiver pixels. 

At BLUSE, a middle ground has been sought: when a signal of interest is detected in a coherent beam, a so-called ``stamp" file is saved, along with the standard total-power hits files. These stamp files contain an upchannelised raw voltage swatch of time-frequency data (encompassing the detected signal) from each antenna; an example is displayed in Figure~\ref{fig:stamps}. Such stamp files allow, for example, re-beamforming anywhere in the primary field of view over the saved frequency range and permitting repeated further analysis. Furthermore, storage requirements are  much easier to manage. The rate at which raw upchannelised data are committed to long-term storage can be controlled by setting the ``dial" of how many stamp files to save per recording (in addition to the SNR threshold described in Section~\ref{sec:search}). For regions of the band where RFI is particularly prevalent, setting a maximum threshold keeps this rate manageable.

In addition to the stamp and hit files, several other data products are recorded. At a user-specified interval  filterbank-formatted HDF5\footnote{\href{https://www.hdfgroup.org/solutions/hdf5/}{https://www.hdfgroup.org/solutions/hdf5/}} files are saved for the incoherent sum of all antennas as well as each coherent beam. These are useful for system validation and diagnostics. Finally, for the rare occasions on which BLUSE is used for direct primary observing, all the raw F-engine voltages are saved for later re-analysis in full.

\subsection{Storage and archiving}
\label{sec:storage}

The bulk of the survey data products are stored and processed \textit{in situ} on the BLUSE computing cluster. During the course of a subarray's lifetime, survey products are written to temporary volumes on each processing node. These consist of magnetic disks in RAID 0 for write speed, minimising the time spent on data transfer so as not to hold up subsequent processing cycles. During the time between active subarrays (when no multicast subscriptions are made) these data are copied to storage nodes (see Section~\ref{sec:hardware}) for long-term storage in Gluster\footnote{\href{https://www.gluster.org/}{https://www.gluster.org/}} volumes backed by RAID 6 disk arrays\footnote{RAID 6 arrays can perform read and write operations on all virtual disks in the presence of any two concurrent physical disk failures}.

\section{Automation} 
\label{sec:automation}

MeerKAT offers a number of different subarray configurations (including simultaneous independent subarrays, each comprised of a subset of the available antennas). Primary observers can select, for example, the observing band (UHF, L, or S0 to S4), the number of coarse channels (1k, 4k or 32k) and the antenna allocation. Observers also specify the primary pointings and calibrator sources. BLUSE is designed to operate commensally alongside primary observations conducted with all possible subarray configurations. 

Automating commensal observing under such heterogeneous conditions requires a versatile and flexible control system (see Table~\ref{tab:automation} and Figure~\ref{fig:system}). Pipeline instances must be allocated to subarrays (on configuration) from a central pool, to which they are returned when a subarray is deconfigured and processing is completed. MeerKAT's network architecture also places constraints on how instances may be allocated: instances connected to a particular leaf switch may only subscribe to consecutive multicast groups corresponding with 1/4th of the full band.
Note that some processing nodes host single pipeline instances, while others host two (see Table~\ref{tab:hardware}). Metadata associated with a particular subarray must be confined and delivered to the corresponding participant processing nodes.


\subsection{State machine architecture}
\label{sec:state-architecture}

A set of state machines, two per subarray, is used to ensure that resources are correctly allocated, metadata are correctly delivered, and recording is correctly synchronised across participating processing instances. 

\begin{figure}
    \centering
    \includegraphics[trim={12pt 0 17pt 0}, clip, width=\columnwidth]{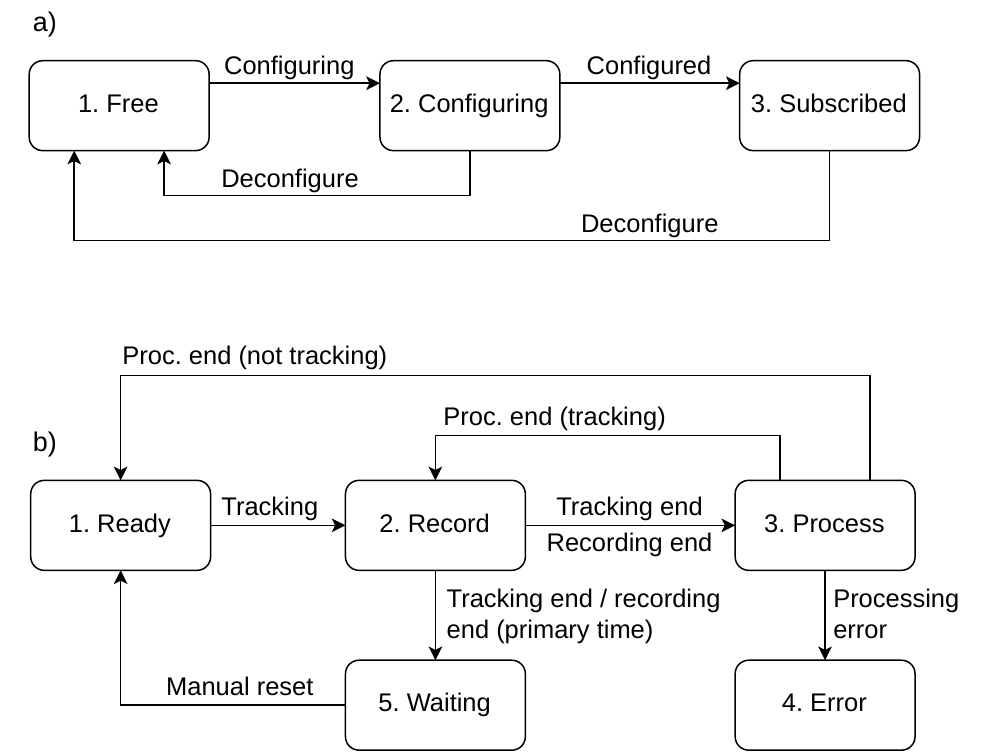}
    \caption{The two different state machines that operate for each possible subarray. The FreeSubscribed states are described in (a) and the RecProc states in (b).}
    \label{fig:state_diagrams}
\end{figure}

The \texttt{FreeSubscribed} states (Figure~\ref{fig:state_diagrams}, top) are used to allocate and de-allocate available processing instances, removing and returning them to the pool of free instances as needed. The \texttt{RecProc} states (Figure~\ref{fig:state_diagrams}, bottom) access the pool of instances that have been allocated to a particular subarray (and which have subscribed to that subarray's multicast streams). 

\begin{figure*}
    \centering
    \includegraphics[width=0.8\textwidth]{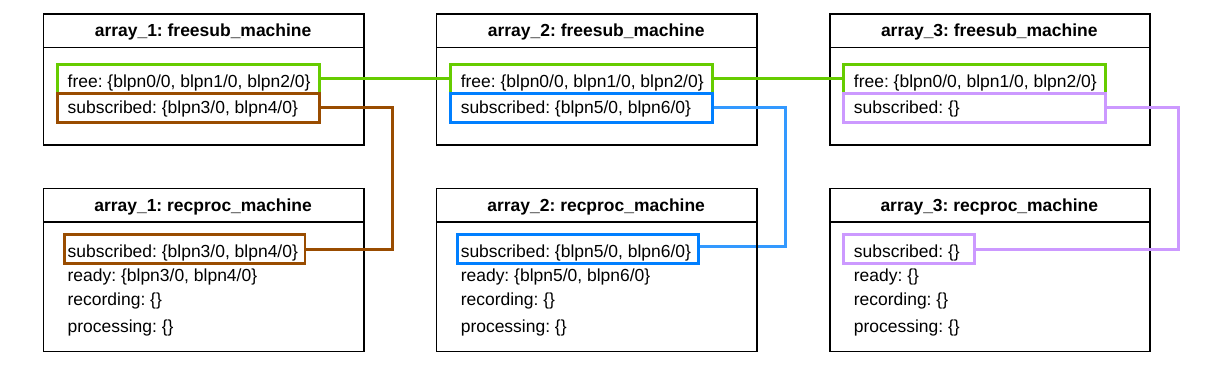}
    \caption{A simplified example of how the coordinator allocates individual pipeline instances to different independent active subarrays.}
    \label{fig:instance_allocation}
\end{figure*}

Figure~\ref{fig:instance_allocation} illustrates this procedure in a simplified example: The boxes and lines in different colours indicate the sets of instances that are shared between state machines. There is only one \texttt{free} set, which is shared by all \texttt{freesubscribed} machines. Each subarray's \texttt{freesubscribed} and \texttt{recproc} machines share a subscribed set. State changes are resolved in sequence, so each machine has its own turn to move instances between sets.

\subsection{State preservation}
\label{sec:state-saving}

Each subarray's \texttt{recproc} and \texttt{freesubscribed} states are saved in {\sc Redis}\footnote{Redis is an in-memory database that persists on disk. \href{https://github.com/redis/redis}{https://github.com/redis/redis}} as $JSON$-formatted dictionaries. This takes place every time a state change occurs. On startup, for each subarray, the coordinator checks to see if state data have been saved previously. If so, it initialises each state machine with the appropriate state and instance allocation, recreating the last known configuration when the coordinator was last active. This allows the coordinator to recover from interruptions and for modifications to be made in the midst of an active subarray.

\subsection{Pipeline control across hosts}
\label{sec:pipeline-control}

\begin{figure*}
    \centering
    \includegraphics[width=\textwidth]{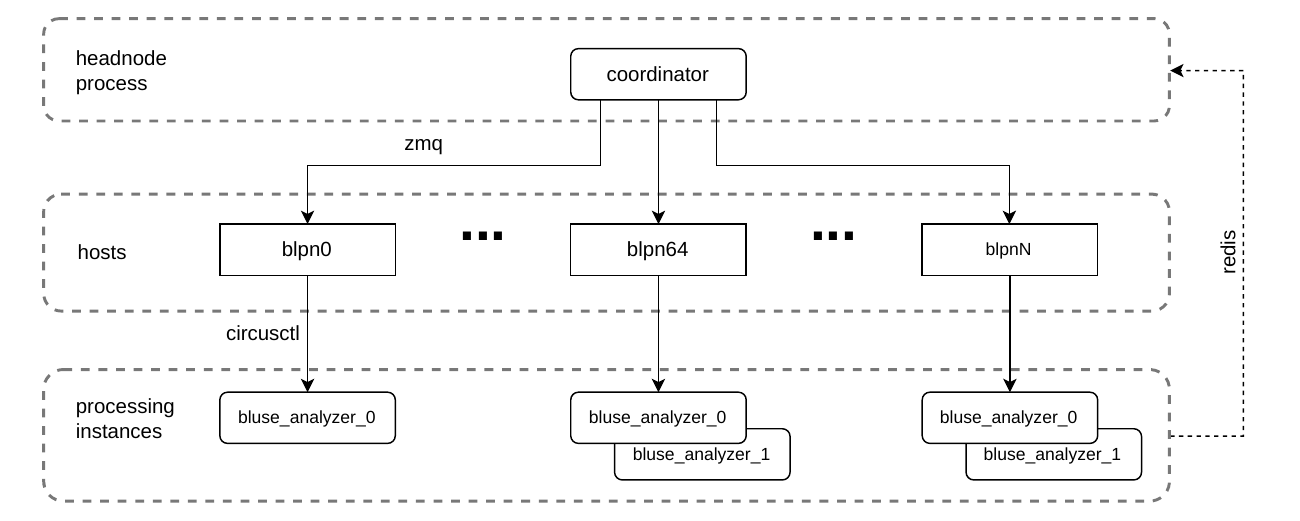}
    \caption{The mechanism by which the coordinator controls processing across processing instances.}
    \label{fig:coordinator}
\end{figure*}

A central coordinator process\footnote{\href{https://github.com/UCBerkeleySETI/commensal-automator}{https://github.com/UCBerkeleySETI/commensal-automator}} controls each individual processing pipeline instance via \texttt{circus} and ZeroMQ; see Fig. ~\ref{fig:coordinator}. \texttt{circus} is widely used to control all ancillary processes and manage log rotation on all hosts in the BLUSE cluster.

\subsection{Automation processes}
\label{sec:automation-processes}

\begin{table*}
\centering
\caption{Automation processes.}
\label{tab:automation}
\begin{tabular}{p{0.18\linewidth} p{0.76\textwidth}}
\toprule
\textbf{Process} & \textbf{Description} \\
\midrule
\texttt{katcp\_interface} & BLUSE proxy \\
\texttt{katportal\_interface} & Metadata retrieval \\
\texttt{coordinator}$^{a}$ & Automates commensal observing \\
\texttt{bfr5\_generator}$^{b}$ & Creates beamformer recipe files for each observation. \\
\texttt{slack\_proxy} & Delivers Slack messages \\
\texttt{targets\_minimal}$^{c}$ & Selects targets for observation and analysis \\
\texttt{bluse\_raw\_watcher}$^{d}$ & Maintains awareness of the contents of the NVMe buffers. One per processing node. \\
\texttt{bluse\_analyzer} & Controls processing. One per instance, running on the processing nodes. \\
\texttt{bluse\_gateway}$^{e}$ & Hashpipe-Redis gateway process \\
\bottomrule
\end{tabular}
\\[2pt]
\raggedright
\footnotesize
$^{a}$ \url{https://github.com/UCBerkeleySETI/commensal-automator}\\
$^{b}$ \url{https://github.com/david-macmahon/BluseBeamformerRecipes.jl}\\
$^{c}$ \url{https://github.com/danielczech/targets-minimal/}\\
$^{d}$ \url{https://github.com/david-macmahon/BluseRawWatch.jl}\\
$^{e}$ \url{https://github.com/david-macmahon/rb-hashpipe}
\end{table*}

The \texttt{katcp\_interface} responds to requests by the MeerKAT Control and Monitoring subsystem (CAM) when BLUSE is included in a subarray, indicating that BLUSE is ready to receive data. Once included, the \texttt{katportal\_interface} connects to CAM via websockets to receive metadata, such as the current pointing, the centre frequency, whether or not the antennas are on target, etc. These metadata are continuously and instantaneously updated. The \texttt{coordinator} maintains state machines for each active subarray, and allocates and controls individual recording and processing instances across the cluster. Targets are selected and delivered by the \texttt{targets\_minimal} process (see \ref{sec:targets}). The head node also hosts the \texttt{slack\_proxy} and the \texttt{bfr5\_generator}. The former provides a bridge (via {\sc Redis}) for all the other processes to communicate with Slack\footnote{\href{https://slack.com/}{https://slack.com/}} for reporting purposes. The latter generates \texttt{bfr5} files for downstream use by the beamformer and technosignature search software (i.e. seticore). 
 
Each processing node also hosts the \texttt{bluse\_raw\_watcher}, the \texttt{bluse\_gateway} and the \texttt{bluse\_analyzer} (one per processing instance; see Table~\ref{tab:automation} for further details). 

\subsection{Target selection} 
\label{sec:targets}
Target selection is handled by a separate process, \texttt{targets\_minimal}, which runs on the head node. The coordinator sends requests via {\sc Redis}, providing it with the current observing band and the coordinates of the current primary pointing. It estimates the current primary beam width at half maximum power, and retrieves a list of targets for observation that fall within the beam. These targets are ranked in order from most to least desirable, based on prior observations in the different observing bands, and distance. 

When a target is observed and successfully processed in a particular band, it is assigned a simple additive score $S'_{band}$ for that particular band, calculated as follows:

\begin{equation}
S'_{band} = S_{band} + t \times b \times n
\end{equation}

where $S_{band}$ is the previous score, $t$ is the duration of the current observation, $b$ is the number of subband segments, and $n$ is the number of antennas. 

When new targets are selected, they are ranked (in ascending order) first by their scores in the current band, then by the sum of their scores in the other bands, and then finally by distance. The observing ranking is then:

\begin{enumerate}
    \item Targets that have been observed the least in the current band.
    \item Targets that have been observed the least in all other bands combined.
    \item Distance. 
\end{enumerate}

The main target list contains approximately 32 million stars (all-sky) derived from Gaia DR2 and other sources. See \cite{Czech_2021} for a detailed description of how these sources were selected. In addition, a provisional extended `semi-Exotica' sample of 2 million other objects (e.g., galaxies, AGNs, stars) created as an expanded counterpart to the Breakthrough Listen exotica catalogue \citep{Lacki_2021} has been integrated into BLUSE's target list (Lacki et al., in prep).

\subsection{Monitoring} 
\label{sec:monitoring}

System performance and observing progress are monitored in several different ways. Many processes report their status (and provide updates) via Slack. Grafana and Prometheus are used to monitor and visualise many aspects of system health, including (for example) NVMe write rates, CPU usage, etc. Some of the automation processes also transmit annotations to Grafana, which are then displayed on the resultant plots (for example, the time at which recording started for a particular observation). Daily observing summaries are automatically emailed to system maintainers.

\section{Hardware}
\label{sec:hardware}

\begin{figure}
    \centering
    \includegraphics[width=\columnwidth, trim={0 0 0 -5pt}]{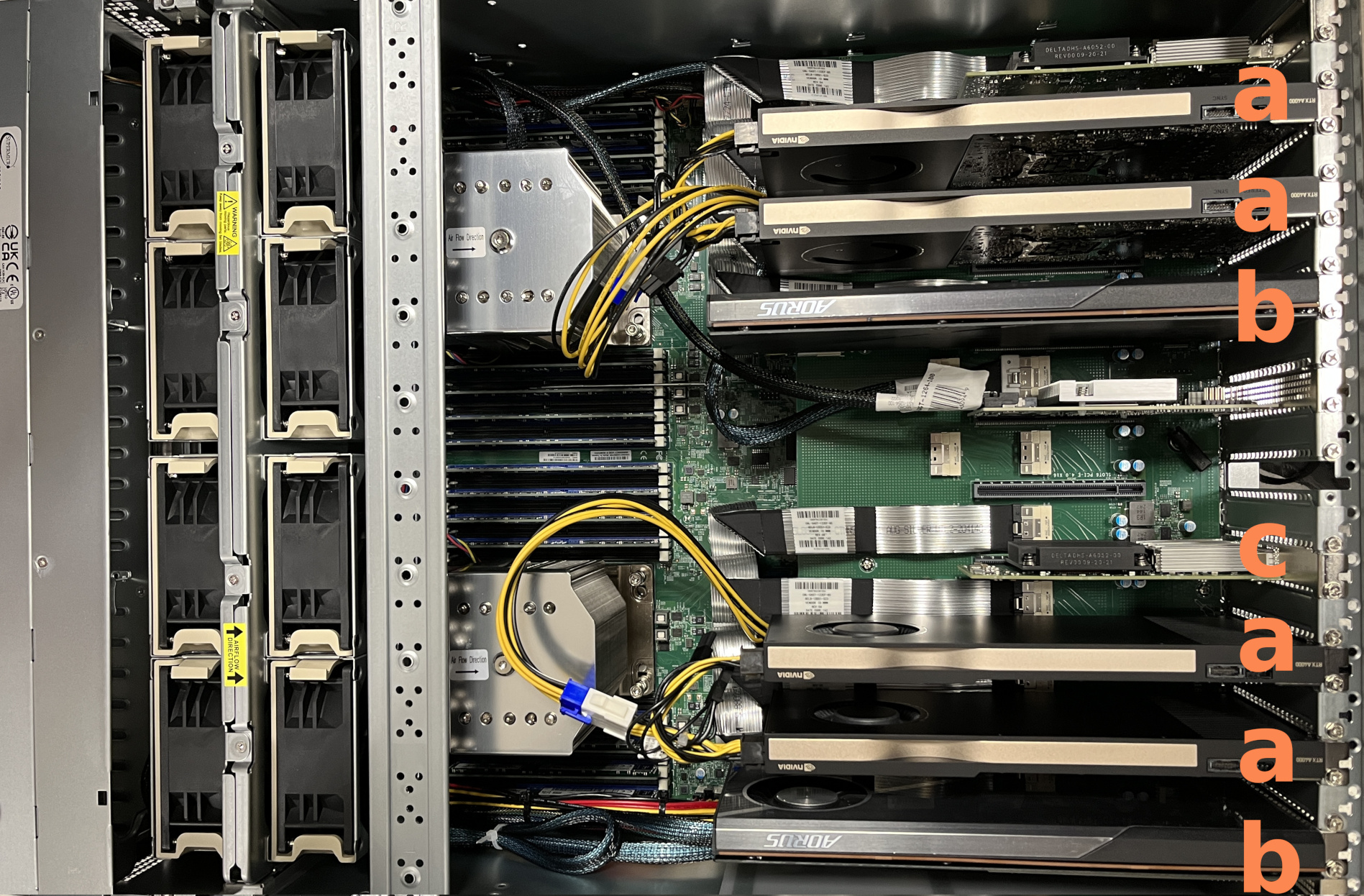}
    \caption{AMD-based processing node, showing the four RTX A4000 Ampere GPUs (a), the two NVMe carrier cards (b), and the NIC (c).}
    \label{fig:amd-node}
\end{figure}

\begin{figure}
    \centering
    \includegraphics[width=\columnwidth]{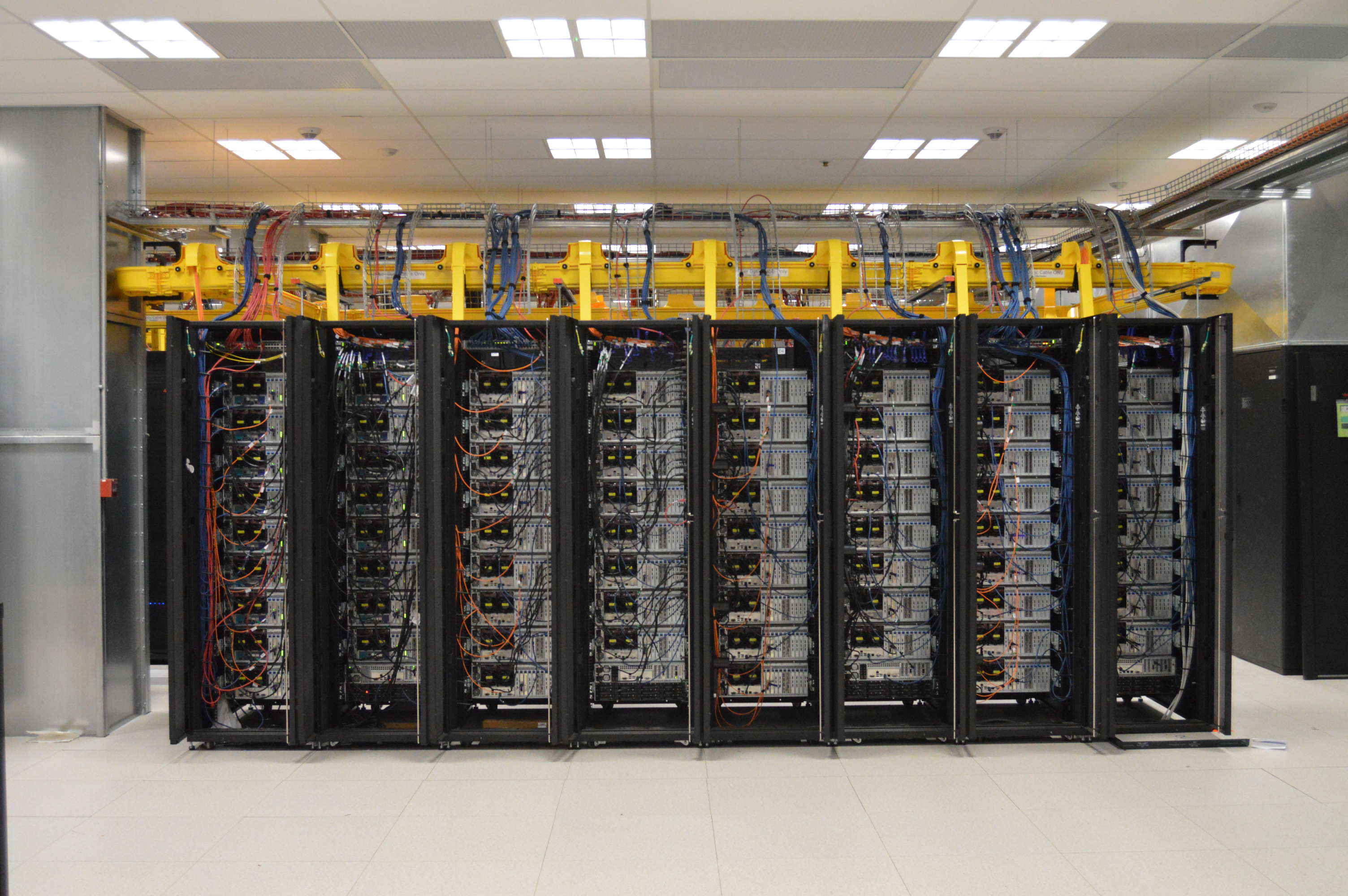}
    \caption{8 of the 16 racks occupied by BLUSE in the KAPB. Servers physically closer to the 40GbE switches use copper interconnect (black), while those further away use fibre (orange).}
    \label{fig:racks}
\end{figure}

BLUSE is implemented on a large computing cluster, hosted in 16 racks in the Karoo Array Processor Building (KAPB), hosted by the South African Radio Astronomy Observatory (SARAO). Table~\ref{tab:hardware} provides a detailed breakdown of the current fleet of servers. Figure~\ref{fig:amd-node} illustrates a typical AMD-based processing node, and Figure~\ref{fig:racks} shows 8 of the 16 racks in-situ in the KAPB.

\begin{table*}
\centering
\caption{Current BLUSE hardware co-located with MeerKAT.}
\footnotesize
\begin{tabular}{p{0.2\textwidth} p{0.1\textwidth} p{0.65\textwidth}}
\toprule
\textbf{Node} & \textbf{Quantity} & \textbf{Description} \\
\midrule
Processing (AMD) & 37 & 32 active, 4 hot spares, 1 experimental node. 2x AMD EPYC 7413, 512GB RAM, 4x RTX A4000 each (active nodes), 2x RTX 3090 (spare nodes), 16x 1TB HDDs, 8x 1TB NVMe drives, 2x 100GbE ConnectX 5 NIC. \\
Processing (Intel) & 68 & 64 active, 4 hot spares. 2x Intel Xeon 4208, 96GB RAM, 1x RTX 2080Ti, 6x 8TB HDDs, 4x 512GB NVMe drives, 1x 40GbE ConnectX 5 NIC each. \\
Storage (version a) & 4 & 2x Intel Xeon 4208, 96GB RAM, 36x 8TB HDDs, 50GbE ConnectX 5 NIC \\
Storage (version b) & 4 & 2x Intel Xeon 4309Y, 192GB RAM, 36x 16TB HDDs, 50GbE ConnectX 5 NIC \\
Head node & 1 & 2x Intel Xeon E5-2620, 128GB RAM 4x 960 GB SSDs, 2x 600GB SSDs, \\
\bottomrule
\end{tabular}
\label{tab:hardware}
\end{table*}

Several options are under consideration for future potential hardware upgrades, some of which would be accompanied by architectural changes to the BLUSE pipeline (see Section~\ref{sec:conclusion}).

\section{Validation} 
\label{sec:validation}

In order to validate the system's features, functionality and performance, a suitable series of test observations was carried out. The Voyager spacecraft has been used as a technosignature test source in prior SETI surveys \citep{Welch_2009, Tremblay_2024}, but cannot be used here since MeerKAT does not currently host X-band (8--12\,GHz) receivers. Instead, we observed the James Webb Space Telescope's (\textit{JWST}) S-band telemetry downlink  with MeerKAT's S-band receivers. \textit{JWST} is a good narrowband technosignature test source as it moves relatively slowly in right ascension and declination, remaining well within the primary field of view over a 290 second recording segment. 

To conduct an end-to-end test of the entire pipeline, we observed a fixed pointing in right ascension and declination in the path of \textit{JWST}, recording its short transit across a small portion of the primary field of view. We observed using MeerKAT's S0 band in 4k mode, and recorded the coarse channel data (raw voltages) as well as the standard data products described in Section~\ref{sec:products}. Preserving the coarse channel data allowed us to form as many arbitrarily placed coherent beams as desired, offline after the observation had taken place.

\begin{figure}
    \centering
    \includegraphics[trim={25pt 0 17pt 55pt}, clip, width=\columnwidth]{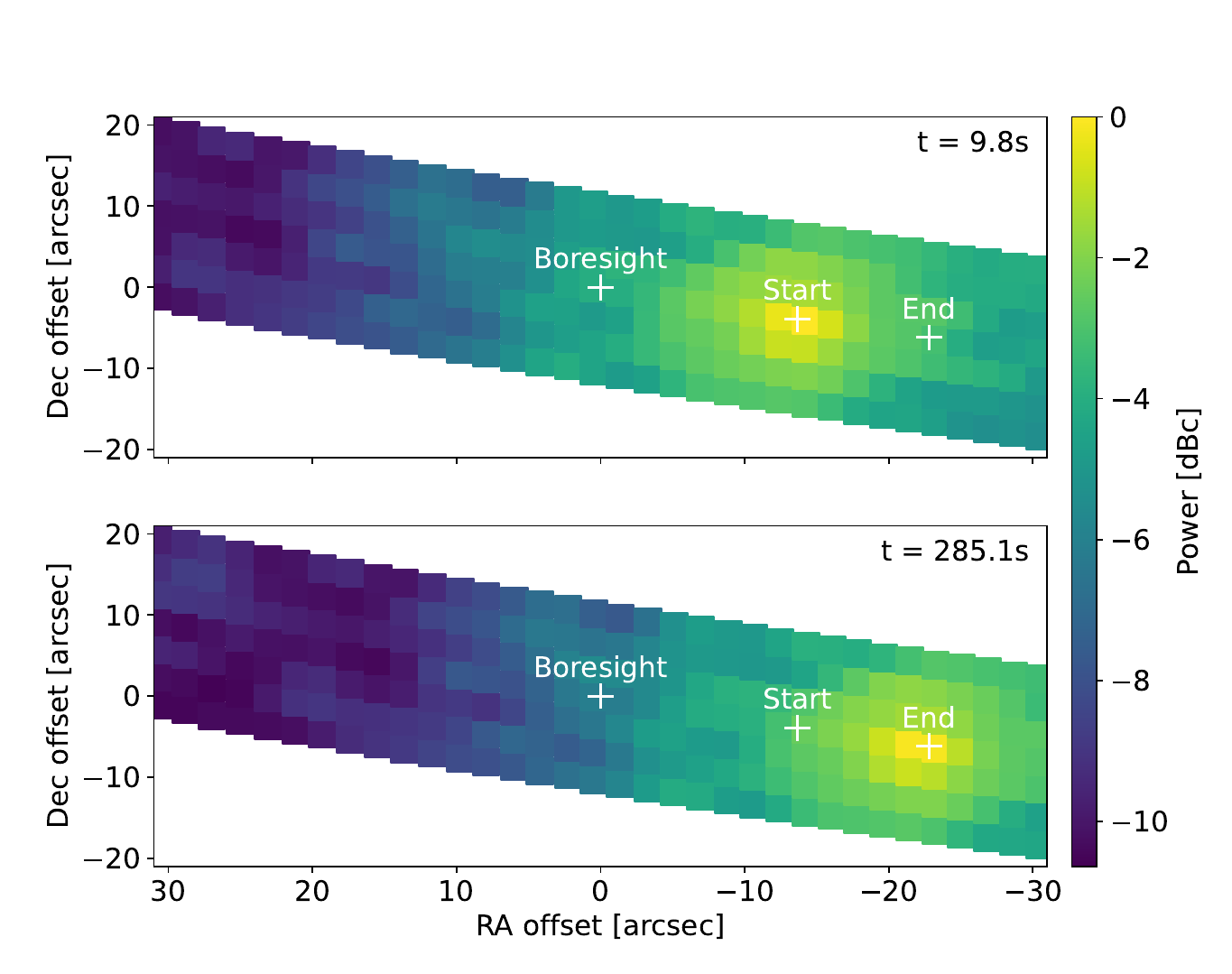}
    \caption{Tiled set of synthesized beams at fixed right ascension and declination at the start and end of the recording. Each beam's power is represented by an individual square marker. For visual purposes, the markers are representative of the beam power but not beam size or shape.  The markers labelled ``Start" and ``End" represent the expected locations of \textit{JWST} at the start and end of the 290 second recording.}
    \label{fig:beam-pattern}
\end{figure}

\begin{figure}
    \centering
    \includegraphics[width=\columnwidth]{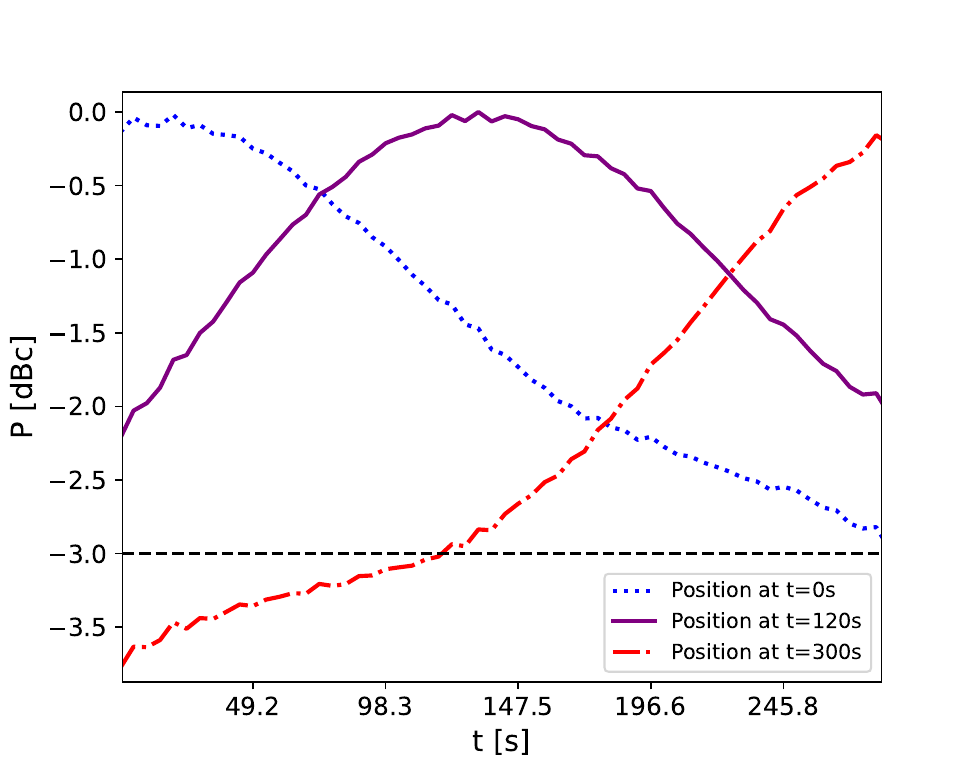}
    \caption{Power over time of different synthesized beams placed at fixed sky coordinates on the path of \textit{JWST's} transit, corresponding with the position of \textit{JWST} at t=0s, t=120s and t=300s.}
    \label{fig:beam-power}
\end{figure}

\begin{figure*}
    \centering
    \includegraphics[width=\textwidth]{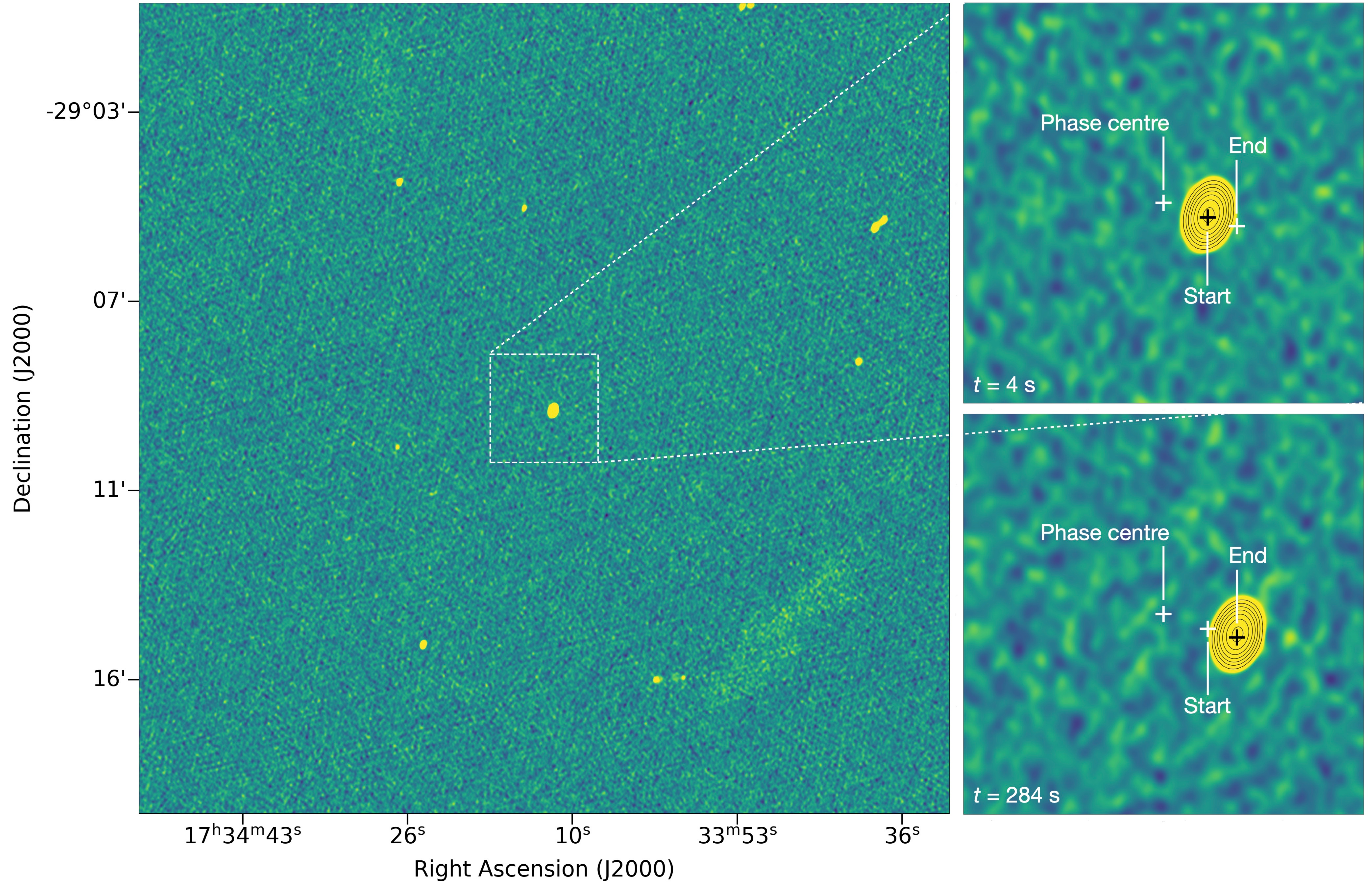}
    \caption{{\bf Main panel:} A 0.3~$\times$~0.3 deg$^{2}$ region of a single 8 second snapshot image formed from the MeerKAT correlator data. The bright central source is the \textit{JWST}. {\bf Upper right panel:} A 1.8~$\times$~1.8 arcmin$^{2}$ zoom of the main panel, showing the position of \textit{JWST} relative to the correlator phase centre, and its expected start and end position for the data discussed in Section \ref{sec:validation}. {\bf Lower right panel:} As per the upper right panel, but for the 35th correlator integration interval of the data. Since the correlator integration time is 8 seconds, these two frames are the closest to the start and end of the 290 second recording segment that was processed by the BLUSE pipeline.}
    \label{fig:jwst-image}
\end{figure*}

Figure~\ref{fig:beam-pattern} illustrates \textit{JWST's} transit during the 290 second recording by means of a set of tiled coherent beams (synthesized offline using the preserved raw voltage data) and placed at fixed sky coordinates. These beams correspond with Figure~\ref{fig:jwst-image}, which shows two frames from a time series of interferometric images with 8~second cadence, produced using data from MeerKAT's correlator. Figure~\ref{fig:beam-power} illustrates the change in power over time of several coherent beams present in Figure~\ref{fig:beam-pattern}.

We also compared the coherent and incoherent beams for this observation, following the procedure described by \cite{Tremblay_2024} to calculate an efficiency ratio as follows:

\begin{equation}
\eta = \frac{P_{coherent}}{P_{incoherent}} \times \frac{1}{N_{ants}}
\end{equation}

where $P_{coherent}$ and $P_{incoherent}$ are normalised as described by \cite{Tremblay_2022}. This yields a ratio $P_{coherent}/P_{incoherent}$ of approximately 52.63, yielding an efficiency of 84.9\% given that there were 62 antennas in the subarray at the time of the observation. Small phase calibration errors and other effects cause the observed efficiency to drop below the theoretical maximum, and the result of 84.9\% is expected.

\cite{rajwade2022} cite a beamforming efficiency (using a different method) of 0.92-0.96, albeit at L-band. Using their method, our efficiency result is approximately 0.92.

\begin{figure*}
    \centering
    \includegraphics[width=\textwidth]{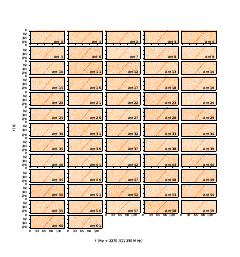}
    \caption{The contents of a stamp file saved automatically in response to a detection of \textit{JWST's} S-band telemetry downlink in a synthesized beam. The same time-frequency region of the upchannelised voltage data from each participating antenna is extracted and saved. In this observation, 62 antennas were available. This plot shows power for both polarisations, summed.}
    \label{fig:stamps}
\end{figure*}

Figure~\ref{fig:stamps} provides an example of the contents of a stamp file that the pipeline saved in response to a detection of \textit{JWST's} telemetry downlink. The time-frequency region surrounding the detection is saved for every participating antenna. In each subplot, the signal from \textit{JWST} is clearly visible as expected.

\section{Observing progress} 
\label{sec:progress}

BLUSE has been operating autonomously since mid-2022.  By developing improvements to the automation software, we have increased reliability and overall observing rate: in the first year of operations, from mid-2022 to mid-2023, we processed approximately 73,500 coherent beams, while from mid-2023 to mid-2024 we processed approximately 386,200, a five-fold increase. From mid-2024 to mid-2025 we processed approximately 436,700 coherent beams.

During the three years from 2023 to 2026, we processed approximately 1.5 million coherent beams in total. Of these 1.5 million, approximately 1.2 million were viable for technosignature searching (the remainder were too short in duration, less than 150s). Of these 1.2 million, approximately 360,000 unique objects were observed: that is to say, many objects were re-observed several times when the same primary pointing coordinates were revisited. Given improvements to BLUSE, observing efficiency has increased, and we estimate that we are now observing around 29,000 globally unique objects per month.

The resulting data products are currently being analysed and will be presented in a series of forthcoming scientific publications, several of which are presently under review. These studies will include investigations of the distribution of detected signals as a function of observing frequency. In addition, we have published initial survey results towards K2-18b in the \textit{Astrophysical Journal} \citep{tremblay_2026}.

\begin{figure*}
    \centering
    \includegraphics[width=\textwidth]{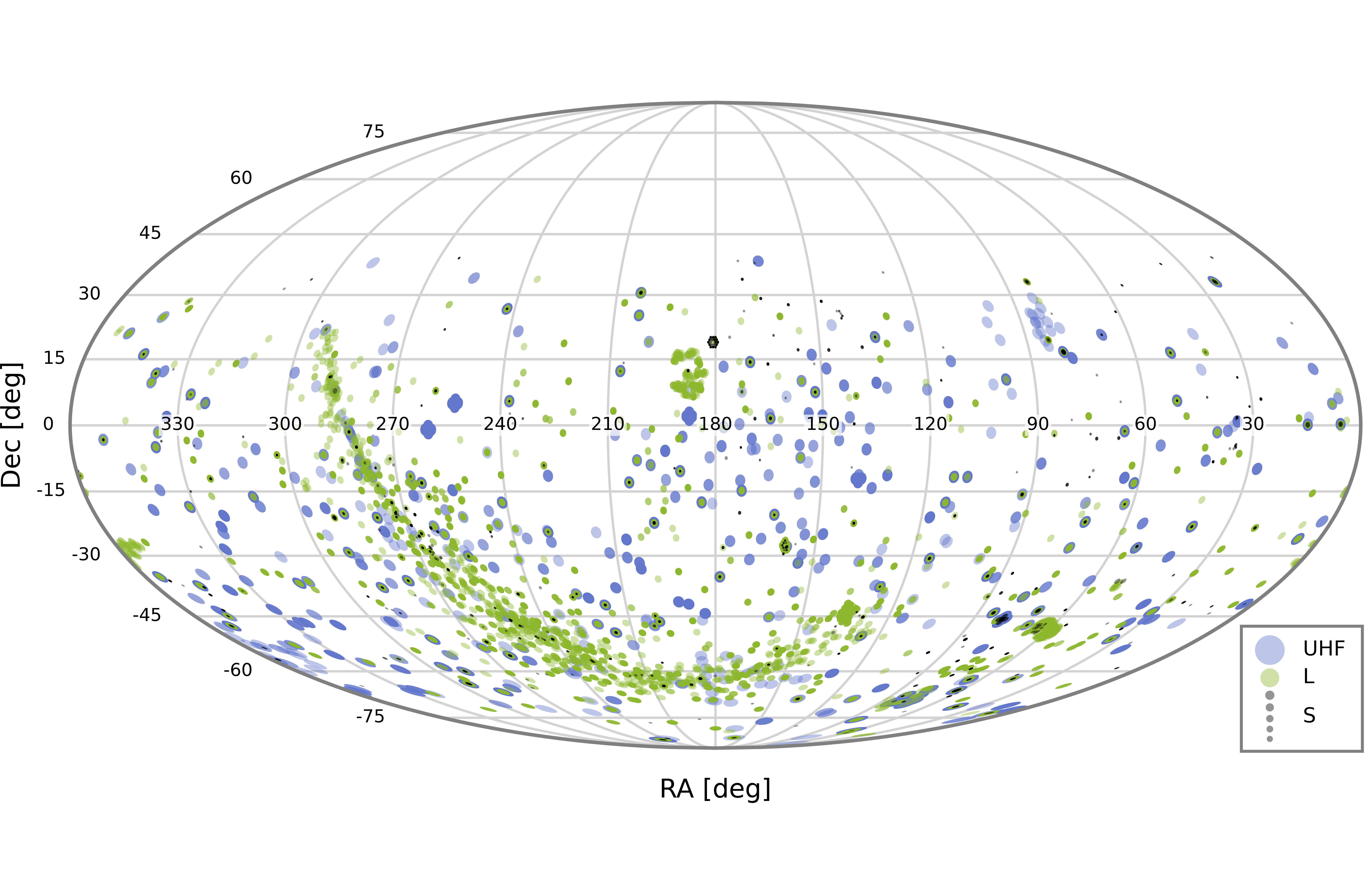}
    \caption{Sky coverage of commensal observing thus far, showing the primary fields of view (to scale) for each processed primary pointing in UHF, L and S-band. Translucency has been applied to each marker, so overlapping fields appear darker (for example where fields have been observed repeatedly). Note the generous density of observations along the galactic plane, over the Virgo cluster and the Euclid Deep Field (South), among others.}
    \label{fig:sky-coverage}
\end{figure*}

Figure~\ref{fig:sky-coverage} illustrates the sky coverage of the commensal survey thus far. Various primary observing campaigns can be discerned in this plot. Along with generous coverage of the Galactic plane, tiling observations of (for example) the Euclid Deep Field (South) and the Virgo Cluster are also apparent.

\section{Conclusions and future work} 
\label{sec:conclusion}

Radio telescope arrays are well suited to large-scale, target-agnostic commensal surveys. MeerKAT offers excellent opportunities for commensal observing, which Breakthrough Listen has taken advantage of with BLUSE, Breakthrough Listen's User Supplied Equipment system. BLUSE is an automated commensal technosignature survey system that ingests the full bandwidth of incoming data produced by the F-engines, forms coherent beams on targets of interest and conducts searches for technosignatures. In this work, we introduce BLUSE's systems from a technical standpoint and provide thorough experimental validation of their features and capabilities. We achieve this by observing the James Webb Space Telescope's S-band telemetry downlink, an excellent technosignature test signal to evaluate the BLUSE system in its entirety, end-to-end. Finally, we discuss observing progress and sky coverage over the past few years of operations, during which BLUSE has processed and analysed approximately 1.2 million individual pointings, nearly all of which were 290s in duration. 

Several exciting endeavours are planned and underway for BLUSE. Besides the standard commensal observing approach described in this work, BLUSE has also been used for and alongside several primary time observations, including a campaign to study the K2-18 system \citep{tremblay_2026} and the interstellar comet 3I/Atlas (\textit{in prep.}). We are also supporting several student projects in the form of pipeline components that are controlled via the \texttt{analyzer} processes. These include a project to perform beam-tiling observations of galaxy clusters when they fall within the primary field of view. 
We plan to explore a variety of different detection algorithms separately and in conjunction with \texttt{seticore}, including e.g. \texttt{bliss}\footnote{\href{https://github.com/n-west/bliss}{https://github.com/n-west/bliss}} and interferometric imaging-based approaches.
Further in the future it would be advantageous to upgrade BLUSE's hardware to take full advantage of available GPUDirect RDMA capabilities, increasing performance by bypassing the CPUs and perhaps the NVMe recording buffers (see for example recent work at the Allen Telescope Array by \cite{Ma_2025}).
The rise of target-agnostic, commensal, array-based technosignature surveys such as BLUSE and COSMIC \citep{Tremblay_2024} heralds a step change in the rate at which searches are conducted. MeerKAT, with its southern sky coverage, sensitivity and RFI-quiet environment, and BLUSE, with its commensal versatility and raw performance, are among the vanguard driving the field forward.

\newpage
\section*{Acknowledgements}

The MeerKAT telescope is operated by the South African Radio Astronomy Observatory, which is a facility of the National Research Foundation, an agency of the Department of Science, Technology and Innovation. 

As part of the Breakthrough Listen project, BLUSE is sponsored by the Breakthrough Initiatives, affiliated with the Breakthrough Prize Foundation.

\section*{Data availability}

The data underlying this article will be shared on request to the corresponding author.

\bibliography{bibliography}{}
\bibliographystyle{mnras}

\bsp	
\label{lastpage}
\end{document}